\documentclass[12pt]{article}
\usepackage{amssymb,latexsym}
\usepackage{epsfig}
\usepackage{psfig}
\usepackage{graphicx}

\begin{document}

\newcommand{\vm}{v_{\rm max}}

\title{Self-organized patterns and traffic flow in colonies 
of organisms: 
from bacteria and social insects to vertebrates{\footnote{A longer version of this article will be published elsewhere}}}

\author{Debashish Chowdhury$^{1}$, Katsuhiro Nishinari$^{2}$, and 
Andreas Schadschneider$^3$\\[0.5cm]
$^{1}$ Department of Physics\\ 
Indian Institute of Technology\\
Kanpur 208016, India\\[0.3cm]
$^2$ Department of Applied Mathematics and Informatics\\
Ryukoku University\\ 
Shiga 520-2194, Japan\\[0.3cm]
$^3$ Institute for Theoretical Physics\\
Universit\"at zu K\"oln\\ 
50937 K\"oln, Germany}

\maketitle
\abstract{Flocks of birds and schools of fish are familiar examples of 
spatial patterns formed by living organisms. In contrast to the patterns 
on the skins of, say, zebra and giraffe, the patterns of our interest 
are {\it transient} although different patterns change over different 
time scales. The aesthetic beauty of these patterns have attracted the 
attentions of poets and philosophers for centuries. Scientists from 
various disciplines, however, are in search of common underlying 
principles that give rise to the transient patterns in colonies of  
organisms. Such patterns are observed not only in colonies of organisms 
as simple as single-cell bacteria, as interesting as social insects like 
ants and termites as well as in colonies of vertebrates as complex as 
birds and fish but also in human societies. In recent years, particularly 
over the last one decade, physicists have utilized the  conceptual 
framework as well as the methodological toolbox of statistical mechanics 
to unravel the mystery of these patterns. In this article we present an 
overview emphasizing the common trends that rely on theoretical modelling 
of these systems using the so-called agent-based Lagrangian approach.}

%%%%%%%%%%%%%%%%%%%%%%%%%%%%%%%%%%%%%%%%%%%%%%%%%%%%%%%%%%%%%%%%%%%%%%%%%%%%%
%%%%%%%%%%%%%%%%%%%%%%%%%%%%%%%%%%%%%%%%%%%%%%%%%%%%%%%%%%%%%%%%%%%%%%%%%%%%%

\section{Introduction} 

In general, pattern is a general term for any recognizable regularity 
in the observed data. By ``spatial'' pattern we mean some kind of 
regularity in the arrangment of the constituents in space \cite{levin}. 
Similarly, when we monitor a time-dependent quantity over a period of 
time, any possible regularity in the temporal variation may be referred 
to as a ''temporal'' pattern \cite{glass,goldbeter,winfree}. Moreover, 
in phenomena where both spatial and temporal regularities occur 
simultaneously, a decoupling of the analysis of spatial and temporal 
patterns may not be possible and one has to deal with a spatio-temporal 
pattern \cite{hess}.  

Exotic patterns observed in living systems have attracted attention 
of physicists for a long time \cite{thomson}. During the development 
of an organism, i.e., during the process of morphogenesis, cells are 
known to form specific patterns in tissues that form parts of specific 
organs or organ systems \cite{turing,meinhardt1,meinhardt2,gierer}. However, 
in this paper we shall consider almost exclusively patterns exhibited 
by aggregates of organisms in their colonies \cite{allee}; schools of 
fish and flocks of migrating birds are, perhaps, the most familiar 
patterns of this type \cite{parrish,ball}.  The cellular patterns in 
tissues, after an initial transient period, remain practically unchanged 
for the remaining life period of the organism. In contrast, patterns 
exhibited by aggregates of organisms are transient; some of these 
aggregate are short-lived whereas others may persist for days. 

Such aggregation is observed in colonies of organisms as simple as 
single-cell bacteria as well as in colonies of complex multi-cellular 
vertebrates. All types of locomotion, i.e., aerial, aquatic and 
terrestrial, of individual organisms can give rise to such aggregation. 
The population of organisms in the aggregate may vary from tens to 
millions. These aggregates come in a wide range of shapes and sizes. 

The reasons for the formation of such aggregates are now quite well 
understood. Individual organisms benefit (e.g., escaping predators) 
from aggregation despite some detrimental effects (e.g., getting 
infected by contagious disease). However, several fundamental 
questions regarding the structure and dynamics of these aggregates 
remain to be answered. For example, (a) what decides the shape of 
an aggregate and how do these form, (b) how does an aggregate maintain 
a shape over a period of time, (c) what triggers the changes of shapes 
of the aggregates, (d) how does fission and fusion of existing aggregates 
take place \cite{gueron1}, etc.

Encouraged by the success of the conceptual framework of non-equilibrium 
statistical mechanics in the study of pattern formation in non-living 
systems \cite{gollub,levin2,hohenberg}, efforts have been made over the 
last decade to understand pattern formation in living systems by applying 
the same conceptual tools. It turns out that the aggregates of organisms 
exhibit richer patterns than those observed in non-living systems; this 
may be due to the fact that the constituent elements (i.e., the individual 
organisms) are living objects with many internal degrees of freedom. 
What makes these living organisms so different from their non-living 
counterparts is that each living object is an autonomous system that 
is capable of taking decision which is normally in its own self-interest. 
Thus, the evolution of the patterns involves a subtle interplay of the 
dynamical response of the individual organisms to their local surroundings 
and the global dynamics at the level of the colonies. 

On the basis of the formation process, these aggregates can be broadly
divided into two classes: (i) aggregates that "self-organize", and
(ii) aggregates that form in response to external cues such as light
or food. There are situations where an external cue nucleates an
aggregate, but the aggregate soon grows in size dwarfing the original
stimulus \cite{camazine}. For example, a small school of fish may 
nucleate around some floating object, but soon the school may grow to 
such a huge size that the original attractant becomes irrelevant.

In the transient patterns exhibited by aggregates of organisms, one 
can identify two different characteristic time scales of dynamics: 
the shorter time scale is associated with the reflex and response of 
individual organisms to their immediate surroundings whereas the 
whole pattern changes on the longer time scale. In those patterns 
where terretrial locomotion drives each individual organisms and the 
pattern consists of long linear stretches, the spatio-temporal 
organization appears very similar to those in vehicular traffic. 
Perhaps, the most familiar examples are traffic of ants and termites 
on trails. In this article we shall also analyse such traffic flows 
from the perspective of statistical physics \cite{css,helbing,nagatani}.

A common feature of most of the patterns considered in this paper is 
that each organism can be represented by a (self-propelled) particle 
and the system, as a whole, may be regarded as a collection of 
interacting (self-propelled) particles driven far from equilibrium 
\cite{schweitzer}. The steady states of such systems are of current 
interest in non-equilibrium statistical mechanics \cite{szia,schutz}. 
The transition from one such steady state to another, with the variation 
of parameters, is analogous to phase transitions exhibited by thermodynamic 
systems in equilibrium. The non-equilibrium phase transitions from one 
dynamical phase to another remain among the most challenging and least 
understood frontiers of statistical physics.

%%%%%%%%%%%%%%%%%%%%%%%%%%%%%%%%%%%%%%%%%%%%%%%%%%%%%%%5
\section{Theoretical approaches for modeling} 
%%%%%%%%%%%%%%%%%%%%%%%%%%%%%%%%%%%%%%%%%%%%%%%%%%%%%%%5

The fundamental question to be addressed by any theory of patterns in 
colonies of organisms is the following: 
how do the {\it individual} decisions and {\it local} interactions of 
the individuals influence the {\it global} structure (shape, size, etc.), 
{\it collective} dynamics and function of the colony of the organisms? 
What is the interplay of deterministic and stochastic dynamics? 

\subsection{Different types of theoretical approaches}

First of all, the theoretical approaches can be broadly divided into 
two categories: 
(I) ``Eulerian'' and (II) ``Lagrangian''. In the {\it Eulerian} models 
individual organisms do not appear explicitly and, instead, one 
considers only the population densities (i.e., number of individual 
organisms per unit area or per unit volume). But, the {\it Lagrangian} 
models  describe the dynamics of the individual organisms explicitly.   
Just as ``microscopic'' models of matter are formulated in terms of 
molecular constituents, the Lagrangian models of pattern are also 
developed in terms of the constituent organisms. Therefore, the 
Lagrangian models are often referred to as ``microscopic'' models. 

In the recent years, it has been emphasized by several groups (see, 
for example, ref.~\cite{flierl}) that although the patterns of the 
colonies are manifest only at the level of the population, the patterns 
are emergent collective properties that are determined by the 
responses of the individuals to their local environments and the 
local interactions among the individual organisms. Therefore, in 
order to gain a deep understanding of the pattern formation process, 
it is essential to investigate the linkages between these two levels 
of biological organization.

Usually, but not necessarily, space and time are treated as continua 
in the Eulerian models and partial differential equations (PDEs) or 
integro-differential equations are written down for the time-dependent 
local collective densities of the organisms 
\cite{okubo,grunbaum,parr}. The Lagrangian models have been formulated 
following both continuum and discrete approaches. In the continuum 
formulation of the Lagrangian models, differential equations describe 
the individual trajectories of the organisms \cite{flierl}. 

For developing Lagrangian model, one must first specify the {\it state} 
of each individual organism. The informations which may be needed for 
the complete specification of the state include, for example, the 
location (position in space whose dimensionality may be one, two or 
three), genotype  and phenotype, ontogenetic status (age, size and 
maturity), physiological status (hunger), behavioral status (motivation), 
etc. \cite{flierl}. In addition, the environmental informations must 
also be provided; these may include physical and chemical features of 
the environment, resources in the environment, etc. The dynamical 
laws governing the time-evolution of the system must predict the 
state of the system at a time $t + \Delta t$, given the corresponding 
state at time $t$. The change of state should reflect the response of 
the system in terms of movement of the individual organisms, their 
mortality, reproduction (and consequent population growth). 

A natural framework for the mathematical formulation of such models is 
the Newton's equations for individual organisms; each organism is 
modelled as a ``particle'' subjected to some ``effective forces'' 
arising out of its interaction with the other organisms in the colony 
\cite{mogil3,gueron2}. 
These forces not only cause their movements but also their alignments. 
In addition, the organisms may also experience viscous drag and some 
random forces (``noise'') that may be caused by the surrounding fluid 
medium. Even in the absence of any direct physical interaction between 
the organisms (other than the ``hard-core'' repulsion) there may be 
some other ``effective interactions'' which are often referred to as 
``social'' interactions. Some of these interactions capture the effects  
of communications via chemical signalling; these include, for example, 
communications among amoeba forming a multicellular slug or those 
between ants on a trail. This type of models are sometimes formulated 
in terms of an effective energy landscape. Each organism executes moves 
in its own energy landscape which, in turn, varies with time because of 
the movement of the other organisms \cite{rauch}. In contrast to the 
forces arising from physical interactions, the social forces do not 
necessarily obey Newton's third law! In this paper we shall present an 
explicit example for such social interactions in the context of the 
pedestrian dynamics.

Most of the recent Lagrangian models, however, have been formulated 
on discretized space and the temporal evolution of the system in 
discrete time steps are prescribed as dynamical update rules using 
the language of cellular automata (CA) \cite{wolfram,chopard} or 
lattice gas (LG) \cite{marro}. Since each of the individual organisms 
may be regarded as an agent, the CA and LG models 
are someties also referred to as agent-based models \cite{pnas}.

One advantage of the continuum models is that all the tools for analytical 
treatment of differential- and integro-differential equations are readily 
available \cite{holmes,mogil2}. However, a continuum formulation is usually 
sensible only for large and {\it dense} aggregates but hard to justify for 
loosely packed aggregates. Moreover, even if a continuum description can be 
justified, what is even harder to justify is the analytical form of the 
inter-organism interactions, which are required for writing down the 
equations of motion for the individual organisms in the Lagrangian approach.
Furthermore, the differential equations often turn out to be too complicated 
to be solved analytically. Numerical solution of these equations require 
discretization of both space and time. Therefore, the alternative discrete 
formulations, based on CA and LG, may be used from the beginning \cite{edelrev}. 

In fact, there are some further advantages in modeling biological systems 
with CA and LG. Biologically, it is quite realistic to think in terms of 
the way each individual organism responds to its local environment and 
the series of actions they perform. The lack of detailed knowledge of these 
behavioral responses is compensated by the rules of CA. Usually, it is much 
easier to devise a reasonable set of logic-based rules, instead of cooking 
up some effective force for dynamical equations, to capture the behaviour 
of living organisms. Moreover, because of the high speed of simulations of 
CA and LG, a wide range of possibilities can be explored which would be 
impossible with more traditional methods based on differential equations. 
Furthermore, it may be possible to derive continuum Eulerian models by 
appropriately coarse-graining agent-based Lagrangian models under some 
justifiable approximations.

The most satisfactory and convenient {\it analytical} approach may be to use 
a hybrid of the Lagragian-Eulerian methods. One can start with a agent-based 
microscopic model following the Lagrangian approach and, then, derive 
corresponding macroscopic Eulerian models from these equations under reasonable 
approximations. It may be possible to solve the approximate Eulerian equations 
using the analytical tools for solving PDEs.

%%%%%%%%%%%%%%%%%%%%%%%%%%%%%%%%%%%%%%%%%%%%%%%%%%%%%%%%%%%%%%%%%
\subsection{Types of ordering in the aggregate} 
%%%%%%%%%%%%%%%%%%%%%%%%%%%%%%%%%%%%%%%%%%%%%%%%%%%%%%%%%%%%%%%%%

The aggregates formed by the organisms can exhibit different types of 
ordered structures depending on the individual and/or collective 
features as well as external environmental conditions. 

Some aggregates are compact Euclidean objects while others exhibit 
fractal structure. Among the compact patterns, several different types 
of ordering have been observed. For example, the positions and 
orientations of the individual organisms may correspond to those 
of non-spherical molecules in a crystal. However, most often, the 
positions of the organisms do not form a regular lattice but they 
are all oriented more or less in the same direction; some of this 
type of structures of the aggregates are analogues of nematic 
liquid crystals \cite{gruler}.
In some other aggregates like, for example, swarms of mosquitoes, 
there is no spatial or orientational ordering inside the aggregate 
although the aggregate persists.

It has been realized for quite some time that, in reality, swarms have 
a finite size. Therefore, attempts should be made to obtain 
{\it localized} patterns \cite{rappel} or propagating {\it bands} 
\cite{mogil2}, rather than propagating surfaces, to capture migrating 
{\it finite} colonies.

%%%%%%%%%%%%%%%%%%%%%%%%%%%%%%%%%%%%%%%%%%%%%%%%%%%%%%%%%%%%%%%%%
\section{Patterns of aggregates of cells and uni-cellular organisms} 
%%%%%%%%%%%%%%%%%%%%%%%%%%%%%%%%%%%%%%%%%%%%%%%%%%%%%%%%%%%%%%%%%

Various species of uni-cellular organisms, e.g., bacteria, amoeba, 
etc., form aggregates with wide varieties of patterns. Moreover, 
some types of cells, which normally form parts of multi-cellular 
organisms, are also capable of forming interesting patterns when 
these are isolated from the organism and grown in cultures. 

%%%%%%%%%%%%%%%%%%%%%%%%%%%%%%%%%%%%%%%%%%%%%%%%%%%%%%%%%%%%%%%%%
\subsection{Patterns of bacterial colonies in biofilms} 
%%%%%%%%%%%%%%%%%%%%%%%%%%%%%%%%%%%%%%%%%%%%%%%%%%%%%%%%%%%%%%%%%

The microbial biofilms pose not only intellectual challenge to 
physicists interested in the patterns of bacterial colonies but 
also of practical interest in microbiology. For example, the 
dental plaques, the bacterial films formed inside water pipes, 
etc. are examples of microbial biofilms. The growth of such 
structures have been invesigated using CA approaches \cite{wimpenny}.

Normally, in the laboratory, bacterial colonies are grown on substrates 
with a high nutrient level and intermediate agar concentration. 
Patterns of the aggregates formed in colonies grown in such comfortable 
conditions are compact. However, harsh conditions for bacterial 
colonies can be created, for example, by using low level of nutrients; 
aggregates of bacteria formed in such environments can be very 
complex and interesting from the point of view of pattern formation 
\cite{benjacob1,benjacob2}. Hydrodynamics is also likely to play some 
role in the growth of the bacterial colonies in wet conditions 
\cite{lega}. 

The colonies of {\it Proteus mirabilis} form circular swarms that 
have a terrace-like structure. Modeling such swarms have been attempted 
so far following, to our knowledge, only Euclidean approach 
\cite{esipov,czirok01}.
Very recently, Indekeu and Giuraniuc \cite{indekeu} have developed 
a CA model of ``nutrient-limited aggregation'' and growth of 
bacterial towers. The three-dimensional system is modeled as a 
simple cubic lattice where each site can be either empty or occupied 
by a bacterium or nutrient or water. Chemotaxis, rather than 
diffusion, drives the growth process.

There are speculations as to the potential applications of the new 
understanding of the mechanisms of pattern formation in bacterial 
colonies to synthesize systems, which are too complicated to produce 
by conventional methods, through self-organization \cite{benjacobpra}.

%%%%%%%%%%%%%%%%%%%%%%%%%%%%%%%%%%%%%%%%%%%%%%%%%%%%%%%%%%%%%%%%%
\subsection{Patterns of aggregates of amoeba; slime molds} 
%%%%%%%%%%%%%%%%%%%%%%%%%%%%%%%%%%%%%%%%%%%%%%%%%%%%%%%%%%%%%%%%%

The slime mold Dictyostelium discoideum can exist in two different 
forms, namely, either as a population of individual amoeba or as 
a multi-cellular organism consisting of thousands of cells 
\cite{loomis,bonner}. These amoeba feed on other bacteria and can 
exist as a well dispersed colony of uncorrelated individual 
organisms when the supply of nutrients is abundant. However, when 
nutrients become scarce, the amoeba begin a collective restructuring 
of the colony through communication via chemical signalling. The 
colony exhibits a sequence of different spatial patterns; the final 
three-dimensional structure is a fruit body that looks similar to 
a small mushroom with very large number of spores. The spores get 
spread out over a large area by wind. When these land in areas with 
sufficiently high supply of nutrients, they give rise to a new well 
dispersed population of amoeba. 

From the perspective of transient patterns, the intermediate stages 
of aggregation of the amoeba are most interesting. The initial 
patterns consist of concentric rings which gradually transform to 
rotating spirals. At a later stage, the slow movements of the 
cells towards the centers of such patterns transform the patterns 
themselves into a system of thin dense streaks.

The aggregation of the amoeba Dictyostelium discoideum was 
modelled long ago (for example, by Keller and Segel \cite{keller}) 
using an Eulerian approach  (see also ref.\cite{rappel2} for the 
recent literature on theoretical models). Kessler and Levin 
\cite{kesslerca} have developed a discrete CA-type model to study 
these spatial patterns. Each cell was represented by a two-state 
automaton (a ``bion'') which is capable of measuring the concentration 
and concentration gradients of cAMP as well as sense the presence 
of nearby bions. 

In the Kessler-Levin model \cite{kesslerca}, initially, a random 
fraction of the sites (typically $5$-$20$ percent) of a square 
lattice are occupied by the bions. In addition, cAMP concentration 
$c$ was assumed to obey a discretized diffusion equation on the 
lattice. Each bion remains in the state $0$ until it detects a local 
concentration above a predetermined threshold. As soon as the local 
cAMP concentration exceeds the threshold, the bion makes a transition 
to the state $1$ and emits an amout $\Delta c$ of cAMP over the next 
$\tau$ time steps. Then the bion remains in a quiescent state $2$ for 
the next $t_R$ time steps before reverting back to the state $0$; in 
the quiescent state the bion remains immune to further excitation. 

The rule for the movement of the individual cells in the Kessler-Levine 
model \cite{kesslerca} is as follows: for a cell located, at a given 
time step $t$, at the lattice site $i,j$, the discretized gradients 
in the concentrations $c$ of cAMP are computed using 
$$(\nabla c)_x = (c_{i+1,j}^{(t)} - c_{i-1,j}^{(t)})/2 $$ 
$$(\nabla c)_y = (c_{i,j+1}^{(t)} - c_{i,j-1}^{(t)})/2 $$ 
If the cell is in the excited state $1$ and if at least one of the two 
gradients exceeds the predetermined threshold, the cell attempts to 
move to the next neighbouring lattice site with the higher concentration 
of cAMP in that direction. If gradients in both the directions exceed 
the threshold, the cell attempts to move in the direction determined 
by the diagonal in between those two directions of increasing cAMP 
concentration. However, the attempt of hopping is successful only if 
the target site is not occupied by any other cell. Each cell can move 
only once in each excitation cycle. Carrying out computer simulations 
of this model, Kessler and Levin \cite{kesslerca} observed the spiral 
spatial patterns characteristic of the aggregation during the formation 
of the multicellular slug. A more detailed CA model, which is intended 
to account for the patterns at different stages of evolution, has 
been developed more recently by Savill and Hogeweg \cite{hogeweg}.

%%%%%%%%%%%%%%%%%%%%%%%%%%%%%%%%%%%%%%%%%%%%%%%%%%%%%%%%%%%%%%%%%
\subsection{Patterns in colonies of myxobacteria} 
%%%%%%%%%%%%%%%%%%%%%%%%%%%%%%%%%%%%%%%%%%%%%%%%%%%%%%%%%%%%%%%%%

Myxobacteria form large clusters and move like a pack of wolves. Each 
individual myxobacterium preys on several other microorganisms. In 
contrast to communication system based on the diffusible morphogens 
in Dictyostelium discoideum, the myxobacteria use an altogether different 
mechanism of communication where cells communicate with each other by 
direct physical contact. Periodic waves of movements, called ripples, 
have been observed in these colonies. However, in spite of superficial 
similarities, there are crucial differences in the properties of these 
ripples and similar patterns observed in cellular slime mold 
Dictyostelium discoideum. In particular, the colliding wave fronts in 
case of Dictyostelium discoideum annihilate each other whereas the waves 
in the colonies of myxobacteria can pass through each other. This 
phenomenon has been reproduced by a model \cite{igoshin}, formulated 
in the spirit of Lagragian approach, where the trajectories of the 
individual myxobacteria are described by a set of Langevin-like 
equations.

%%%%%%%%%%%%%%%%%%%%%%%%%%%%%%%%%%%%%%%%%%%%%%%%%%%%%%%%%%%%%%%%%
\subsection{Pattern of aggregates of fibroblasts} 
%%%%%%%%%%%%%%%%%%%%%%%%%%%%%%%%%%%%%%%%%%%%%%%%%%%%%%%%%%%%%%%%%

Fibroblasts are a special type of cells found in connective tissues. 
In the laboratory these cells can be extracted from their natural 
locations and their in-vitro aggregation can be studied using cultures 
on a petri dish. The two-dimensional colonies of such cells have been 
found to form patches where each patch consists of a single layer of 
hundreds of fibroblasts with a single axis of orientation. The competing 
contiguous patches eventually merge into a large array with one single 
axis of orientation.

A CA model for this pattern formation by fibroblasts was developed by 
Edelstein-Keshet and Ermentrout \cite{edelfibro}.
Each of the lattice sites could be either empty or occupied by a cell.
Each cell is assigned an {\it orientation} and {\it a state of binding}. 
The orientation of a cell can be denoted by an arrow; the direction of 
the arrow determines the direction of the movement of the cell in the 
next time step. The state of binding (i.e., whether or not the cell 
is bound to an aggregate) determines whether or not it is allowed to 
move. However, both the orientation and the state of binding are 
dynamic variables that can change with time.

On a discrete lattice, however, the arrow can point in only a finite 
set of discrete directions. For example, implementing the model on a 
square lattice, Edelshtein-Keshet and Ermentrout \cite{edelfibro} 
allowed the arrow to be pointed towards any of eight neighbouring sites 
surrounding it (i.e., the four nearest-neighbour as well as the four 
next nearest-neibour sites). Consequently, each cell could change its 
orientation by an angle that would be an integral multiple of $45^{\circ}$. 
For the sake of simplicity, only two discrete states of binding are 
assumed; the cell is either bound to an aggregate or unbound (i.e., 
not part of any aggregate). Therefore, the total number of possible 
states of each cell is twelve while that of each lattice site is 
thirteen.

Starting from a homogeneous state with random orientations, the state 
of the system is updated at each discrete time step according to the 
following rules: \\
(i) an unbound cell reorients by one angular unit (e.g., $45^0$ on a 
square lattice) with probability $p_0$, \\
(ii) an unbound cell may reverse its motion with probability $p_R$, \\
(iii) if an unbound cell comes in contact with another cell or group 
of cells, the probabilty of binding and aligning is $p_A$ if the angle 
of contact is small enough; otherwise the approaching cell reverses its 
direction and moves away, \\
(iv) bound cells do not move; the probability that a bound cell detaches 
from a group is $p_D$. 

By carrying out computer simulations, Edelstein-Keshet and Ermentrout 
\cite{edelfibro} demonstrated that this simple model captures the 
essential qualitative features of the aggregation process observed in 
the in-vitro experiments.

%%%%%%%%%%%%%%%%%%%%%%%%%%%%%%%%%%%%%%%%%%%%%%%%%%%%%%%%%%%%%%%%%
\section{Patterns in social insect colonies} 
%%%%%%%%%%%%%%%%%%%%%%%%%%%%%%%%%%%%%%%%%%%%%%%%%%%%%%%%%%%%%%%%%

From now onwards, in this paper we shall study patterns of the 
aggregates formed by multi-cellular organisms. We begin with the 
simpler (and smaller) organisms and, then, consider those of 
organisms with larger sizes and more complex physiology.

Termites, ants, bees and wasps are the most common social insects, 
although the extent of social behavior, as compared to solitary 
life, varies from one sub-species to another \cite{wilson}. The 
ability of the social insect colonies to function without a leader 
has attracted the attention of experts from different disciplines 
\cite{bonabu97,anderson02,huang,bonabu98,theraulaz03,gautrais, 
keshet94,theraulazetal}. Insights gained from the modeling of the 
colonies of such insects are finding important applications  
in computer science (useful optimization and control algorithms) 
\cite{dorigo}, communication engineering \cite{bona00}, artificial 
``swarm intelligence'' \cite{bonabeau} and micro-robotics \cite{krieger} 
as well as in management \cite{meyer}. 

%%%%%%%%%%%%%%%%%%%%%%%%%%%%%%%%%%%%%%%%%%%%%%%%%%%%%%%%%%%%%%%%%
\subsection{Ant-trail formation} 
%%%%%%%%%%%%%%%%%%%%%%%%%%%%%%%%%%%%%%%%%%%%%%%%%%%%%%%%%%%%%%%%%

Ants communicate with each other by dropping a chemical (generically
called {\it pheromone}) on the substrate as they move forward
\cite{wilson,camazine,mikhailov}. Although we cannot smell it, the
trail pheromone sticks to the substrate long enough for the other
following sniffing ants to pick up its smell and follow the trail.
This process is called {\em chemotaxis} \cite{benjacob1}.

Rauch et al.\cite{rauch} developed a continuum model, following a 
hybrid of the Lagrangian and the Eulerian approaches in terms of an 
effective energy landscape. They wrote one set of stochastic 
differential equations for the positions of the ants and another set 
of PDEs for the local densities of pheromone. 

Suppose a set of ``particles'', each of which represents an ant, move 
in a potential field $U[\sigma(x)]$, where the potential at any  
arbitrary location $x$ is determined by the local density $\sigma(x)$ 
of the pheromone field. Consequently, each ``particle'' experiences an  
``inertial''force $\vec{F}(x) = - \nabla U(x)$. Each ``particle'' is 
also assumed to be subjected to a ``frictional force'' where ``friction'' 
merely parametrizes the tendency of an ant to continue in a given 
direction: a smaller ``friction'' implies that the ant's velocity persists 
for a longer time in a given direction. The equation of motion for the 
``particles'' (stochastic differential equation) are assumed to have the 
form \cite{rauch}
\begin{equation}
\ddot{x}  = - \gamma {\dot x} - \nabla U[\sigma(x)] + \eta(t) 
\end{equation}
where $\eta(t)$ is a Gaussian white noise with the statistical properties 
\begin{equation}
\langle \eta(t)\rangle = 0  
\end{equation}
and 
\begin{equation}
\langle\eta(t)\eta(t')\rangle = \frac{1}{\beta} \delta(t-t')  
\end{equation}
The strength $1/\beta$ of the noise determines the degree of determinacy 
with which the particle would follow the gradient of the local potential; 
the larger the value of $\beta$ the stronger is the tendency of the 
particle to follow the potential gradient.

Thus, the movement of an ant may be described as the noisy motion  of a 
particle in an ``energy landscape''. However, this energy landscape is 
not static but evolves in response to the motion of the particle as 
each particle drops pheromone at its own location at a rate $g$ per 
unit time. Assuming that pheromone can diffuse in space with a diffusion 
constant $D$ and evaporate at a rate $\kappa$, the equation governing 
the pheromone field is given by 
\begin{equation}
\frac{\partial \sigma(x)}{\partial t} = D \nabla\sigma(x) + g \rho(x) 
- \kappa \sigma(x)  
\end{equation}
where $\rho(x)$ is the local density of the particles at $x$.
Finally, Rauch et al. assumed that the function $U[\sigma(x)]$ has 
the form 
\begin{equation}
U[\sigma(x)] = - \ln\biggl(1 + \frac{\sigma}{1+\delta \sigma}\biggr)   
\end{equation}
where $1/\delta$ is called the capacity.

Watmough and Edelstein-Keshet \cite{watmough} introduced a CA model 
to study the formation of ant-trail networks by foraging ants. In 
this model, each ant is described by its discrete position and 
velocity on a discrete. The rules for updating the positions and 
velocities of the ants as well as the pheromone contration on the 
trail are as follows:\\
(i) The ants move at a fixed speed; if the ant is a forager and 
not following a trail, its movements are random. \\
(ii) Each ant deposits a trail pheromone at a constant rate as it 
moves.\\
(iii) The trail pheromone also evaporates at a constant rate.\\
(iv) The probability $p_{\ell}(c)$ per unit time that an ant will 
keep following a trail (and not loose the trail) is a function of 
the local pheromone concentration; the function $p_{\ell}(c)$ is 
to be specified separately. \\
(v) When an ant, following a trail, reaches a point of bifurcation 
it chooses one of the two branches where the rule for choosing the 
branch is prescribed in the beginning.\\

Carrying out computer simulations of this CA model, Watmough and 
Edelstein-Keshet \cite{watmough} observed trail patterns that look 
very similar to real ant-trail networks. The formation of human 
trails have some similarities with that of ant-trails; some recent 
models that elucidate the mechanisms of the emergence of human trails 
will be discussed later in this article.

%%%%%%%%%%%%%%%%%%%%%%%%%%%%%%%%%%%%%%%%%%%%%%%%%%%%%%%%%%%%%%%%%
\subsection{Phase transition between disordered and ordered foraging} 
%%%%%%%%%%%%%%%%%%%%%%%%%%%%%%%%%%%%%%%%%%%%%%%%%%%%%%%%%%%%%%%%%

Beekman et al. \cite{ratniks} pointed out close similarities between 
phase transitions in non-living systems and that of foraging behaviour 
on the ant-trail. 

A foraging ant that discovers a food source lays down a pheromone 
trail as it crawls back to its nest. But, the trail would completely 
disappear unless it is reinforced by other ants before the original 
pheromone, a volatile chemical, evaporates away. From the study of 
their model, Beekman et al. \cite{ratniks} found that the two 
important relevant parameters are (a) the total number of 
ants within the colony, and (b) the individual rate at which the 
ants discover food sources. They showed that (i) when the independent 
discoveries of food sources are infrequent, a first order (discontinuous) 
phase transition from a disordered foraging behaviour (i.e., foraging 
without a pheromone trail) to ordered foraging (i.e., trail-based 
foraging) takes place as the size of the colony increses; (ii) this 
transition exhibits hysteresis (i.e., history-dependence): when the 
rate of individual discoveries of food sources decreases the system 
chooses one of the two alternative modes of behaviour- either no trail 
or a well used trail- depending on the initial conditions. In other 
words, when the independent discoveries of food sources become 
infrequent, the colonies find it difficult to start a trail but can 
still sustain an existing trail. 

%%%%%%%%%%%%%%%%%%%%%%%%%%%%%%%%%%%%%%%%%%%%%%%%%%%%%%%%%%%%%%%%%
\subsection{Traffic on ant-trails} 
\label{sub-anttraffic}
%%%%%%%%%%%%%%%%%%%%%%%%%%%%%%%%%%%%%%%%%%%%%%%%%%%%%%%%%%%%%%%%%

In \cite{cgns,ncs,jscn} we have developed particle-hopping models,
formulated in terms of a stochastic cellular automaton (CA) (or,
lattice gas), which may be interpreted as models of uni-directional
and bi-directional traffic flow in an ant-trail. These models are not
intended to address the question of the emergence of the ant-trail
\cite{activewalker}, but focus on the traffic of ants on a trail which
has already been formed. The model generalizes the totally asymmetric
simple exclusion process (TASEP) \cite{derrida1,derrida2,schutz} with
parallel dynamics by taking into account the effect of the pheromone.

\begin{figure}[tb]
\begin{center}
\includegraphics[width=0.65\textwidth]{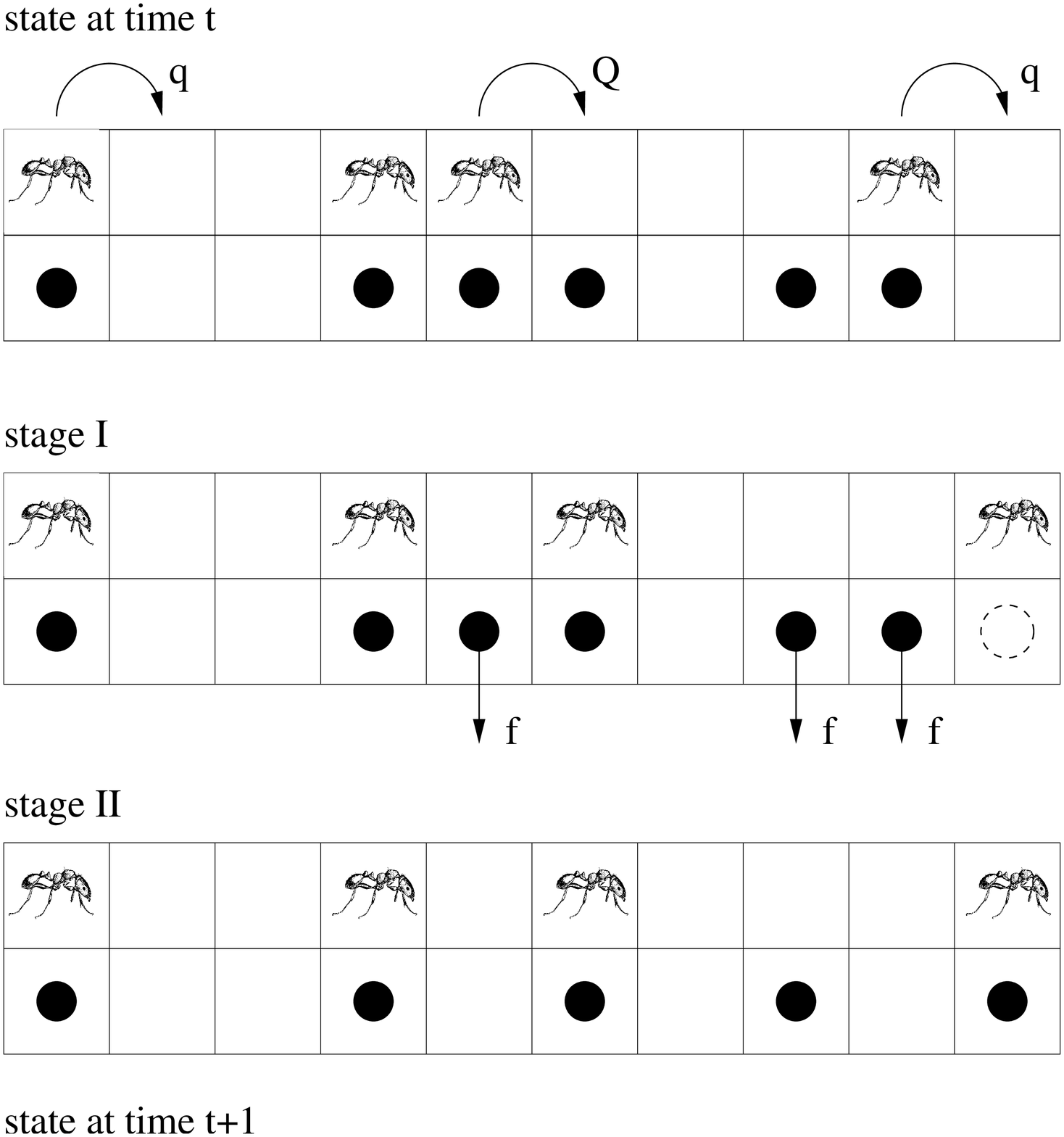}
\end{center}
\caption{
Schematic representation of typical configurations of the
uni-directional ant-traffic model. The symbols $\bullet$ indicate
the presence of pheromone.
This figure also illustrates the update procedure.
Top: Configuration at time $t$, i.e.\ {\it before} {\em stage I}
of the update. The non-vanishing probabilities of forward movement of
the ants are also shown explicitly. Middle: Configuration {\it after}
one possible realisation of {\it stage I}. Two ants have moved compared
to the top part of the figure. The open circle with dashed boundary
indicates the location where pheromone will be dropped by the corresponding
ant at {\em stage II} of the update scheme. Also indicated are the existing
pheromones that may evaporate in {\em stage II} of the updating, together
with the average rate of evaporation.  Bottom: Configuration {\it after}
one possible realization of {\it stage II}. Two drops of pheromones
have evaporated and pheromones have been dropped/reinforced at the
current locations of the ants.
}
\label{fig-modeldef1}
\end{figure}

In our model of uni-directional ant-traffic the ants move according to
a rule which is essentially an extension of the TASEP dynamics.
In addition, a second field is introduced which models
the presence or absence of pheromones (see Fig.~\ref{fig-modeldef1}).
The hopping probability of the ants is now modified by the presence of
pheromones. It is larger if a pheromone is present at the destination site.
Furthermore, the dynamics of the pheromones has to be specified. They
are created by ants and free pheromones evaporate with probability $f$ 
per unit time. Assuming periodic boundary conditions, the state of the 
system is updated at each time step in two stages 
(see Fig.~\ref{fig-modeldef1}). In stage I ants are allowed to move 
while in stage II the pheromones are allowed to evaporate. In each 
stage the {\it stochastic} dynamical rules are applied in parallel to 
all ants and pheromones, respectively.\\

%%%%%%%%%%%%%%%%%%%%%%%%%%%%%%%%%%%%%%%%%%%%%%%%%%%%%%%%%%%%%%%%%%%%
%%%%%%%%%%%%%%%%%%%%%%%%%%%%%%%%%%%%%%%%%%%%%%%%%%%%%%%%%%%%%%%%%%%%

\noindent {\it Stage I: Motion of ants}\\[0.2cm]
\noindent An ant in a site cannot move if the site immediately in front
of it is also occupied by another ant. However, when this site is not
occupied by any other ant, the probability of its forward movement to 
the ant-free site is $Q$ or $q$, depending on whether or not the target 
site contains pheromone. Thus, $q$ (or $Q$) would be the average speed 
of a {\it free} ant in the absence (or presence) of pheromone. To be 
consistent with real ant-trails, we assume $ q < Q$, as presence of 
pheromone increases the average speed.\\

\noindent {\it Stage II: Evaporation of pheromones}\\[0.2cm]
\noindent Trail pheromone is volatile. So, pheromone secreted by an ant
will gradually decay unless reinforced by the following ants. In order to
capture this process, we assume that each site occupied by an ant at the
end of stage I also contains pheromone. On the other hand, pheromone in
any `ant-free' site is allowed to evaporate; this evaporation is also
assumed to be a random process that takes place at an average rate of $f$
per unit time.\\

The total amount of pheromone on the trail can fluctuate although the
total number $N$ of the ants is constant because of the
periodic boundary conditions. In the two special cases $f = 0$ and
$f = 1$ the stationary state of the model becomes identical to that of
the TASEP with hopping probability $Q$ and $q$, respectively.

%%%%%%%%%%%%%%%%%%%%%%%%%%%%%%%%%%%%%%%%%%%%%%%%%%%%%%%%%%%%%%%%%%%%
\begin{figure}[tb]
\begin{center}
\includegraphics[width=0.65\textwidth]{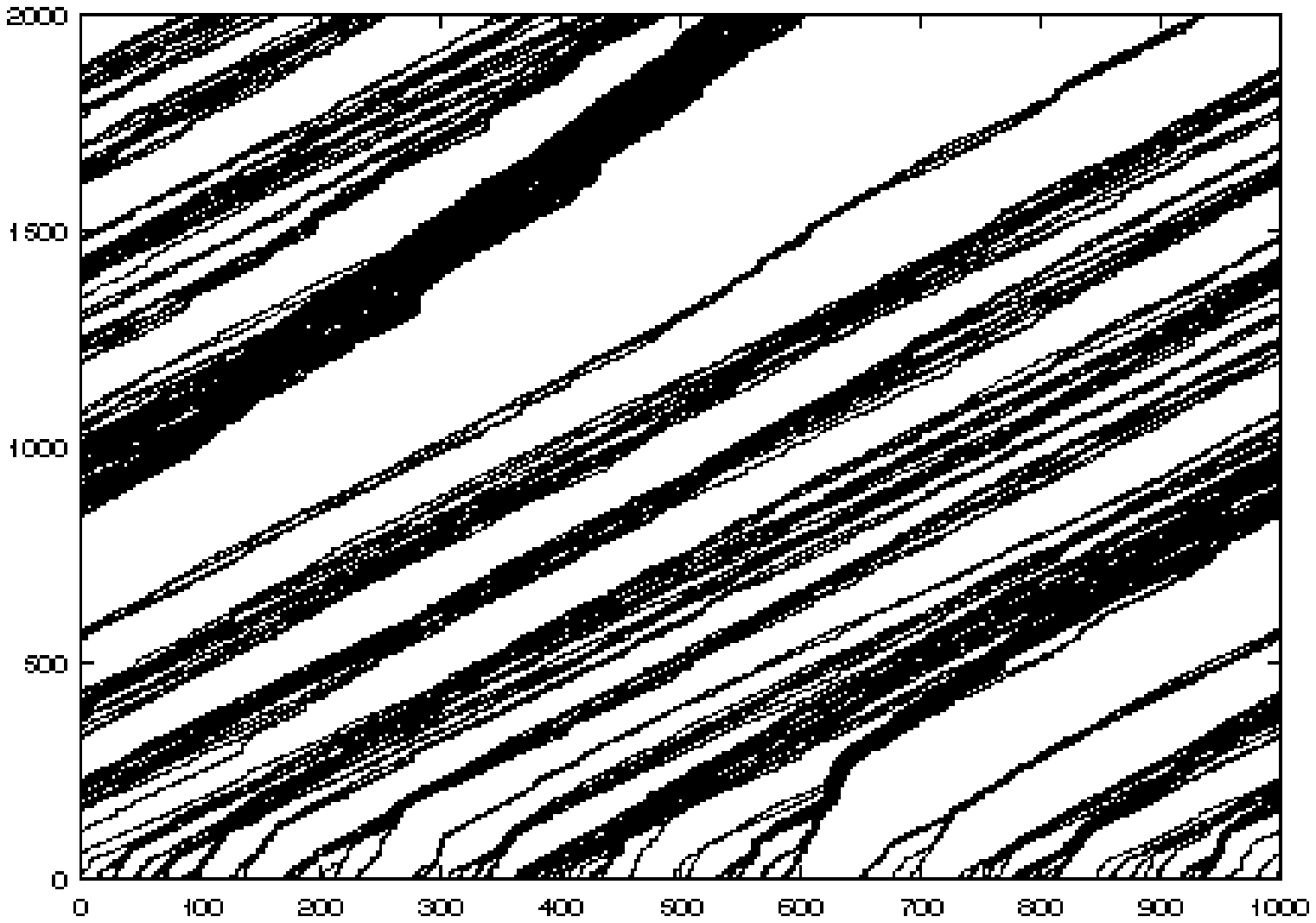}
\includegraphics[width=0.65\textwidth]{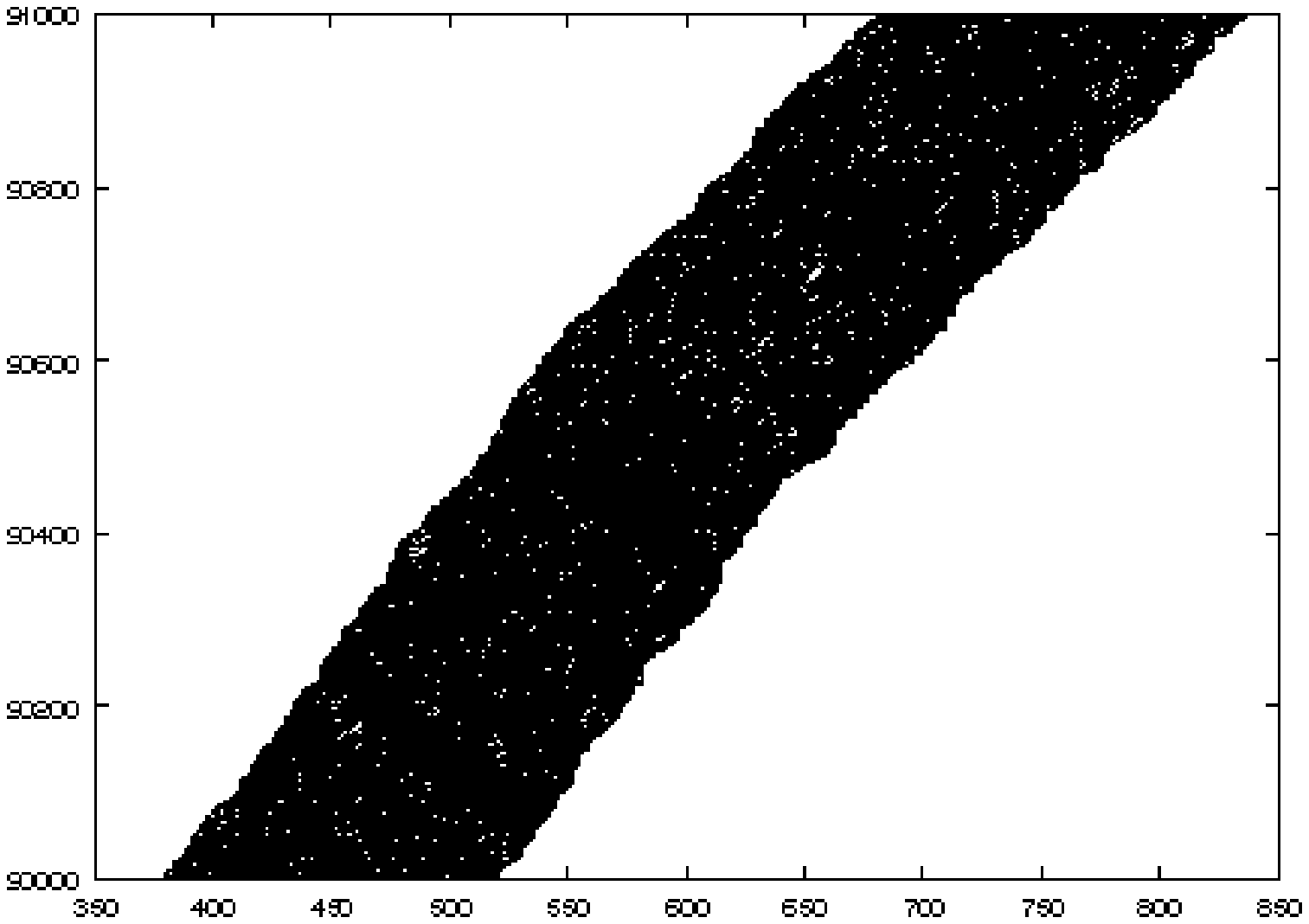}
\end{center}
\caption{
Snapshots of the spatial configurations demonstrating coarsening of 
the clusters of ants.
}
\label{fig-coarse}
\end{figure}
%%%%%%%%%%%%%%%%%%%%%%%%%%%%%%%%%%%%%%%%%%%%%%%%%%%%%%%%%%%%%%%%%%%%

One interesting phenomenon observed in the simulations is coarsening.
At intermediate time usually several non-compact clusters are formed
(Fig.~\ref{fig-coarse}(a)). However, the velocity of a cluster depends
on the distance to the next cluster ahead. Obviously, the probability
that the pheromone created by the last ant of the previous cluster
survives decreases with increasing distance. Therefore clusters with
a small headway move faster than those with a large headway.
This induces a coarsening process such that after long times only
one non-compact cluster survives (Fig.~\ref{fig-coarse}(b)). A similar
behaviour has been observed also in the bus-route model
\cite{busroute1,busroute2}.

In vehicular traffic, usually, the inter-vehicle interactions tend to
hinder each other's motion so that the average speed of the vehicles 
decreases {\it monotonically} with increasing density. In contrast, 
in our model of uni-directional ant-traffic the average speed of the 
ants varies {\it non-monotonically} with their density over a wide 
range of small values of $f$ because of the coupling of their dynamics 
with that of the pheromone. This uncommon variation of the average 
speed gives rise to the unusual dependence of the flux on the density 
of the ants in our uni-directional ant-traffic model. Furthermore, the 
flux is no longer particle-hole symmetric.

It is possible to extend the model of uni-directional ant-traffic 
to a minimal model of bi-directional ant-traffic \cite{jscn}. 
In the models of bi-directional ant-traffic the trail consists of
{\it two} lanes of sites. These two lanes need not be physically 
separate rigid lanes in real space. In the initial configuration, a 
randomly selected subset of the ants move in the clockwise direction 
in one lane while the others move counterclockwise in the other lane. 
The numbers of ants moving in the clockwise direction and 
counterclockwise in their respective lanes are fixed, i.e.\ ants are 
allowed neither to take U-turn\footnote{U-turns of so-called followers 
on pre-existing trails are very rare \cite{beckers}.} nor to change lane.

%%%%%%%%%%%%%%%%%%%%%%%%%%%%%%%%%%%%%%%%%%%%%%%%%
\begin{figure}[tb]
\begin{center}
\includegraphics[width=0.4\textwidth]{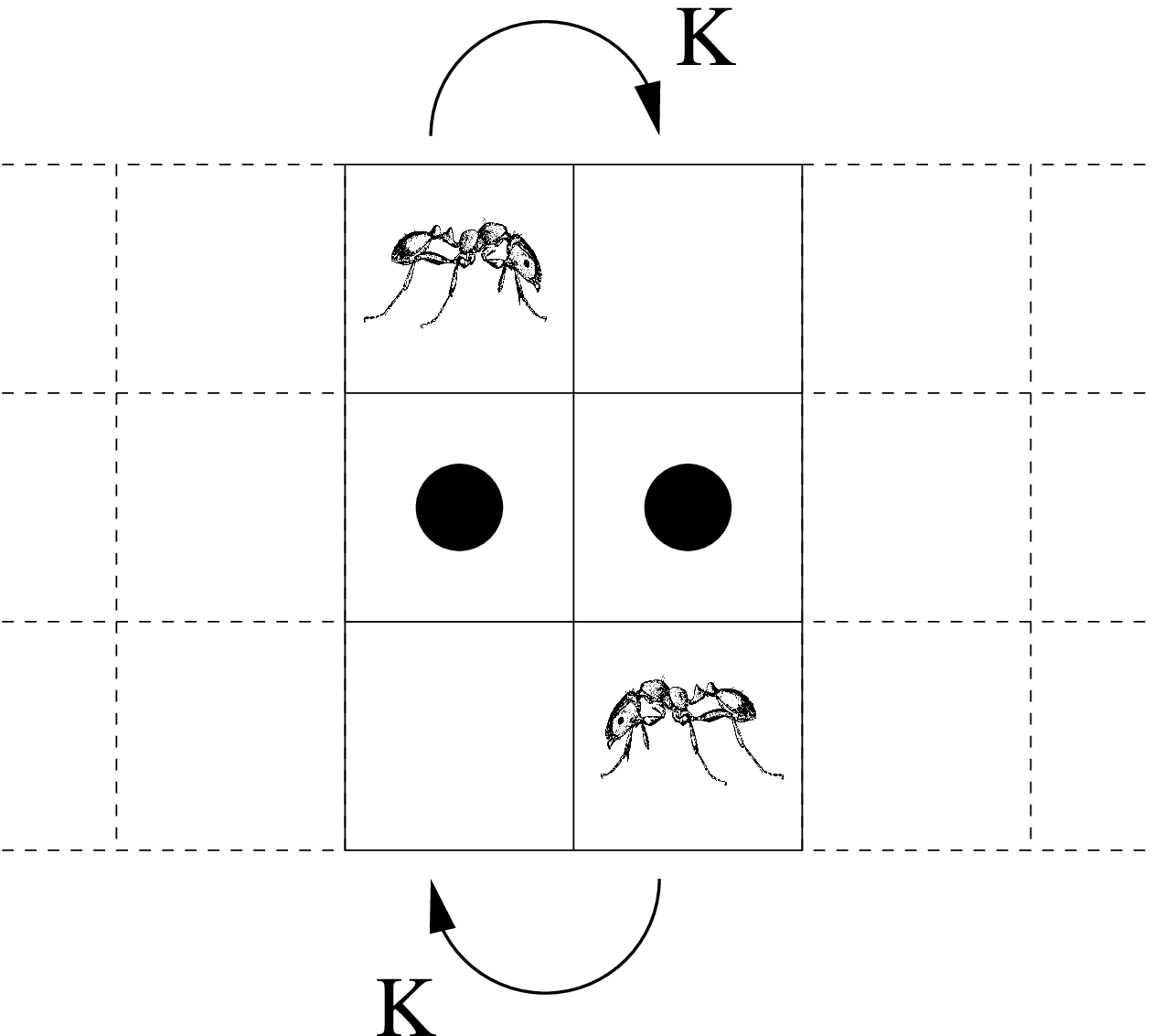}
\end{center}
\caption{A typical head-on encounter of two oppositely moving ants
in the model of {\it bi-directional} ant-traffic.
This is a totally new process which does not have any analog in the
model of uni-directional ant-traffic.
}
\label{fig-modeldef2}
\end{figure}
%%%%%%%%%%%%%%%%%%%%%%%%%%%%%%%%%%%%%%%%%%%%%%%%%

The rules governing the dropping and evaporation of pheromone in the
model of bi-directional ant-traffic are identical to those in the
model of uni-directional traffic. The {\it common} pheromone trail is
created and reinforced by both the outbound and nestbound ants. The
probabilities of forward movement of the ants in the model of 
bi-directional ant-traffic are also natural extensions of the similar 
situations in the uni-directional traffic. When an ant (in either of 
the two lanes) {\it does not} face any other ant approaching it from 
the opposite direction the likelihood of its forward movement onto 
the ant-free site immediately in front of it is $Q$ or $q$, respectively, 
depending on whether or not it finds pheromone ahead. Finally, if an 
ant finds another oncoming ant just in front of it, as shown in 
Fig.~\ref{fig-modeldef2}, it moves forward onto the next site with probability $K$.

Since ants do not segregate in perfectly well defined lanes, head-on
encounters of oppositely moving individuals occur quite often although
the frequency of such encounters and the lane discipline varies from
one species of ants to another. In reality, two ants approaching each
other feel the hindrance, turn by a small angle to avoid head-on
collision \cite{couzin} and, eventually, pass each other.
At first sight, it may appear that the ants in our model follow perfect
lane discipline and, hence, unrealistic. However, that is not true.
The violation of lane discipline and head-on encounters
of oppositely moving ants is captured, effectively, in an indirect
manner by assuming $K < Q$. But, a left-moving (right-moving) ant
{\it cannot} overtake another left-moving (right-moving) ant immediately
in front of it in the same lane. It is worth mentioning that even
in the limit $K = Q$ the traffic dynamics on the two lanes would
remain coupled because the pheromone dropped by the outbound ants also
influence the nestbound ants and vice versa.

\begin{figure}[tb]
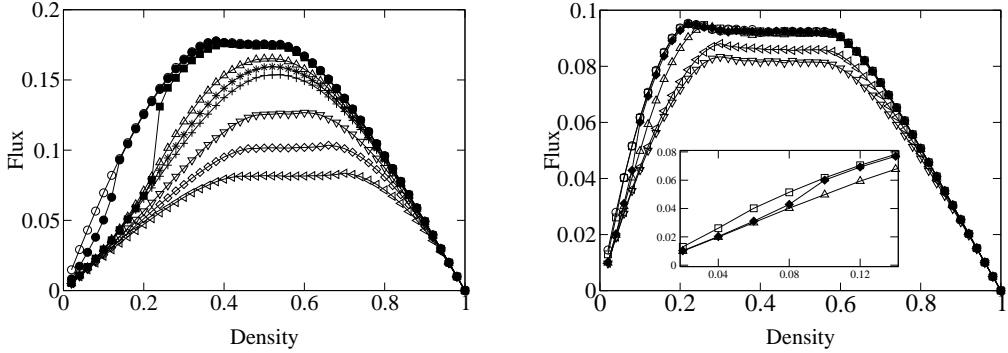

\begin{center}
\ \phantom{a}\\[0.2cm]
\includegraphics[width=0.45\textwidth]{fluxa.eps}
\qquad
\includegraphics[width=0.45\textwidth]{fluxb.eps}
\end{center}
\caption{Fundamental diagrams of the model for bi-directional traffic
for the cases $q<K<Q$ (left) and $K<q<Q$ (right) for several different
values of the pheromone evaporation probability $f$.
{\bf The densities for both directions are identical and therefore
only the graphs for one directions are shown.}
The parameters in the left graph are $Q=0.75, q = 0.25$ and $K=0.5$.
The symbols  $\circ$, $\bullet$, $\blacksquare$,
$\bigtriangleup$, $\ast$, $+$, $\bigtriangledown$, $\Diamond$
and $\triangleleft$ correspond, respectively, to
$f = 0, 0.0005, 0.005,0.05, 0.075,0.10,0.25,0.5$ and $1$.
The parameters in the right graph are $Q=0.75, q = 0.50$ and $K=0.25$.
The symbols $\circ$, $\square$, $\blacklozenge$, $\bigtriangleup$,
$\triangleleft$ and $\bigtriangledown$ correspond, respectively, to
$f = 0, 0.0005, 0.005, 0.05, 0.5$ and $1$.
The inset in the right graph is a magnified re-plot of the same data,
over a narrow range of density, to emphasize
the fact that the unusual trend of variation of flux with density in
this case is similar to that observed in the case $q<K<Q$ (left).
The lines are merely guides to the eye. In all cases curves plotted
with filled symbols exhibit non-monotonic behaviour in the speed-density
relation.
}
\label{fig-flux}
\end{figure}

Fig.~\ref{fig-flux} shows fundamental diagrams for the two relevant
cases $q<K<Q$ and $K<q<Q$ and different values of the evaporation
probability $f$ for equal densities on both lanes.
In both cases the unusual behaviour related to
a non-monotonic variation of the average speed with density
as in the uni-directional model can be observed \cite{jscn}.

%%%%%%%%%%%%%%%%%%%%%%%%%%%%%%%%%%%%%%%%%%%%%%%%%
%%%%%%%%%%%%%%%%%%%%%%%%%%%%%%%%%%%%%%%%%%%%%%%%%

An additional feature of the fundamental diagram in the bi-directional 
ant-traffic model is the occurrence of a plateau region. This plateau 
formation is more pronounced in the case $K<q<Q$ than for $q<K<Q$ since 
they appear for all values of $f$. Similar plateaus have been observed 
earlier \cite{janowsky,tripathy} in models related to vehicular traffic 
where randomly placed bottlenecks slow down the traffic in certain 
locations along the route.

The experimental data available at present \cite{burd1,burd2} are not 
accurate enough to test the predictions mentioned above. More accurate 
measurements, using novel methodologies are in progress \cite{burd3}.

%%%%%%%%%%%%%%%%%%%%%%%%%%%%%%%%%%%%%%%%%%%%%%%%%%%%%%%%%%%%%%%%%
\section{Patterns in the colonies of vertebrates} 
%%%%%%%%%%%%%%%%%%%%%%%%%%%%%%%%%%%%%%%%%%%%%%%%%%%%%%%%%%%%%%%%%

Migrating fish schools and bird flocks have one common feature that 
both of these correspond to a non-vanishing average linear drift 
velocity. In theoretical models a bird or a fish can be represented 
by a {\it polar} self-propelled particle. The nature of the dynamical 
phases and phase transitions of both polar and {\it apolar} self-propelled 
particles have been investigated extensively in the literature over 
the last decade 
\cite{czirokrev,vicetal,csahok,czirok,shimoyama,busse,albano,toner,vic2,mogil1,couzinjtb}.

Ramaswamy and collaborators \cite{sriram} have studied the hydrodynamic 
fluctuations of liquid-crystal-like ordered dynamical phases of 
self-propelled apolar particles. Carrying out linear stability analysis, 
they have not only predicted certain long-wavelength instabilities 
but also indicated the possibility of novel propagating modes which, 
in principle, may be observed in experiments with real or artificial 
self-propelled particles. 

%%%%%%%%%%%%%%%%%%%%%%%%%%%%%%%%%%%%%%%%%%%%%%%%%%%%%%%%%%%%%%%%%
%\subsection{Schools of fish} 
%%%%%%%%%%%%%%%%%%%%%%%%%%%%%%%%%%%%%%%%%%%%%%%%%%%%%%%%%%%%%%%%%

The structure and function of schools of fish have attracted attention 
for the last few dacades \cite{partridge,breder}. But, serious efforts 
have been only over the last few years \cite{reuter,krause} in the 
understanding the mechanism of their formation through self-organization 
in terms of quantitative models. For example, Niwa \cite{niwa} has 
developed a model of fish schooling following the Lagrangian approach. 
However, the equations describing the movements of the individual fishes 
are written in terms of continuous space and time; in fact, these 
equations are very similar to Langevin equations for Brownian particles 
subjected to not only intertial forces but also viscous drag and random 
noise. 

St\"ocker \cite{stocker} has developed a CA model for tuna school 
formation. However, for the sake of simplicity, we outline here the 
main idead behind the first CA model of fish schooling, developed by 
Huth and Wissel \cite{huth}. Each fish is characterized by its 
position and velocity vectors. Suppose, $r_{ij}$ denotes the 
magnitude of the separation between the fish labelled by the integer 
indices $i$ and $j$. In order to decide the position and velocity of 
the fish $i$, one needs to draw three imaginary speheres of radii 
$r_1$, $r_2$ and $r_3$ ($r_1 < r_2 < r_3$) around it. If the fish $j$ 
is located within the smallest sphere, then it would have a repulsive 
effect on the fish $i$ such that the fish $i$ will have a tendency to 
swim away in a direction perpendicular to the direction of the velocity 
of the fish $j$. On the other hand, if the fish $j$ is located anywhere 
within the distance $r_1 < r_{ij} < r_2$, the fish $i$ will tend to 
move parallel to the velocity of the fish $j$. In case $r_2 < r_{ij} < r_3$, 
the fish $i$ will be attracted towards the fish $j$. Finally, if the 
fish $j$ is located outside the largest sphere of radius $r_3$, i.e., 
$r_{ij} > r_3$, it will have no influence on the movement of the fish $i$. 
Huth and Wissel \cite{huth} also introduced rules for combining the 
influences of more than one fish within the spheres of influence.

%%%%%%%%%%%%%%%%%%%%%%%%%%%%%%%%%%%%%%%%%%%%%%%%%%%%%%%%%%%%%%%%%
%\subsection{Flocks of birds} 
%%%%%%%%%%%%%%%%%%%%%%%%%%%%%%%%%%%%%%%%%%%%%%%%%%%%%%%%%%%%%%%%%

%%%%%%%%%%%%%%%%%%%%%%%%%%%%%%%%%%%%%%%%%%%%%%%%%%%%%%%%%%%%%%%%%
\section{Human traffic on trails}
%%%%%%%%%%%%%%%%%%%%%%%%%%%%%%%%%%%%%%%%%%%%%%%%%%%%%%%%%%%%%%%%%

Various kinds of pattern formation can also be observed in human 
societies, especially in pedestrian dynamics \cite{pedebook}. As we 
will see, the human ``intelligence'' plays only a minor role. 
Instead, the observed effects can be understood as simply collective 
phenomena in systems of interacting particles. In fact, some effects 
(like lane formation) also appear in true physical systems \cite{loewen}.
Before we present a CA model that reproduces the essentials of 
pedestrian dynamics we list some of the observed collective phenomena.

\subsection{Collective phenomena and pattern formation}
\label{sec_coll}

One of the reasons why the investigation of pedestrian dynamics is
attractive for physicists is that many interesting collective effects 
and self-organization phenomena can be observed \cite{pedebook,helbing}.

{\bf Jamming}: 
At large densities various kinds of jamming phenomena occur, 
typically at bottlenecks like doors or narrowing corridors. 
This kind of clogging effect does not depend strongly on the 
microscopic dynamics of the particles.
Other types of jamming occur in the case of counterflow where
two groups of pedestrians mutually block each other. This happens
typically at high densities and when it is not possible to turn around
and move back, e.g.\ when the flow of people is large.

\begin{figure}[h]
  \begin{center}
    \includegraphics[width=0.7\textwidth]{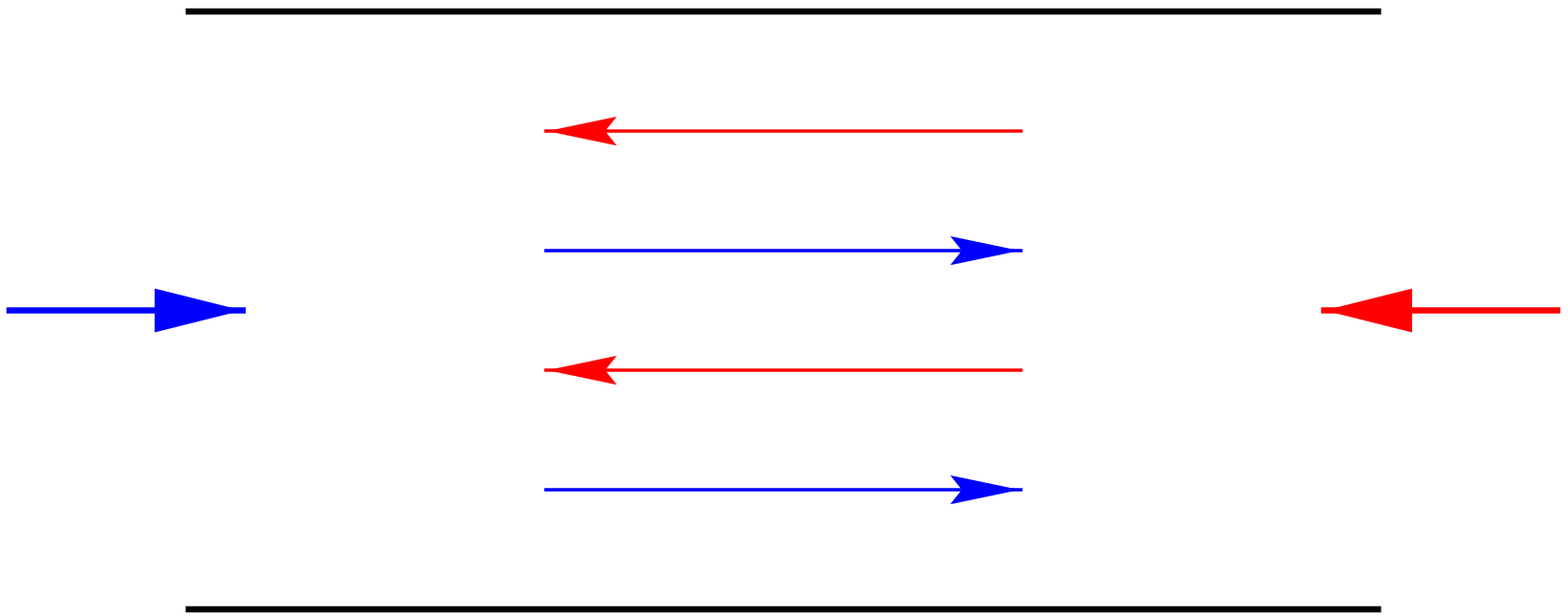}
    \caption{Lane formation in counterflow in a narrow corridor.}
\label{fig_lane}
  \end{center}
\end{figure}

{\bf Lane formation}: In counterflow with groups of
people moving in opposite directions, a kind of spontaneous symmetry
breaking occurs (see Fig.~\ref{fig_lane}). 
The motion of the pedestrians can self-organize into dynamically varying 
lanes where people move in just one direction \cite{social}. 
Thus, strong interactions with oncoming pedestrians are reduced and 
a higher walking speed is possible.

%%%%%%%%%%%%%%%%%%%%%%%%%%%%%%%%%%%%%%%%%%%%%%%%%%%%%%%%%%%%%%%%%
%%%%%%%%%%%%%%%%%%%%%%%%%%%%%%%%%%%%%%%%%%%%%%%%%%%%%%%%%%%%%%%%%

\begin{figure}[h]
  \begin{center}
    \includegraphics[width=0.8\textwidth]{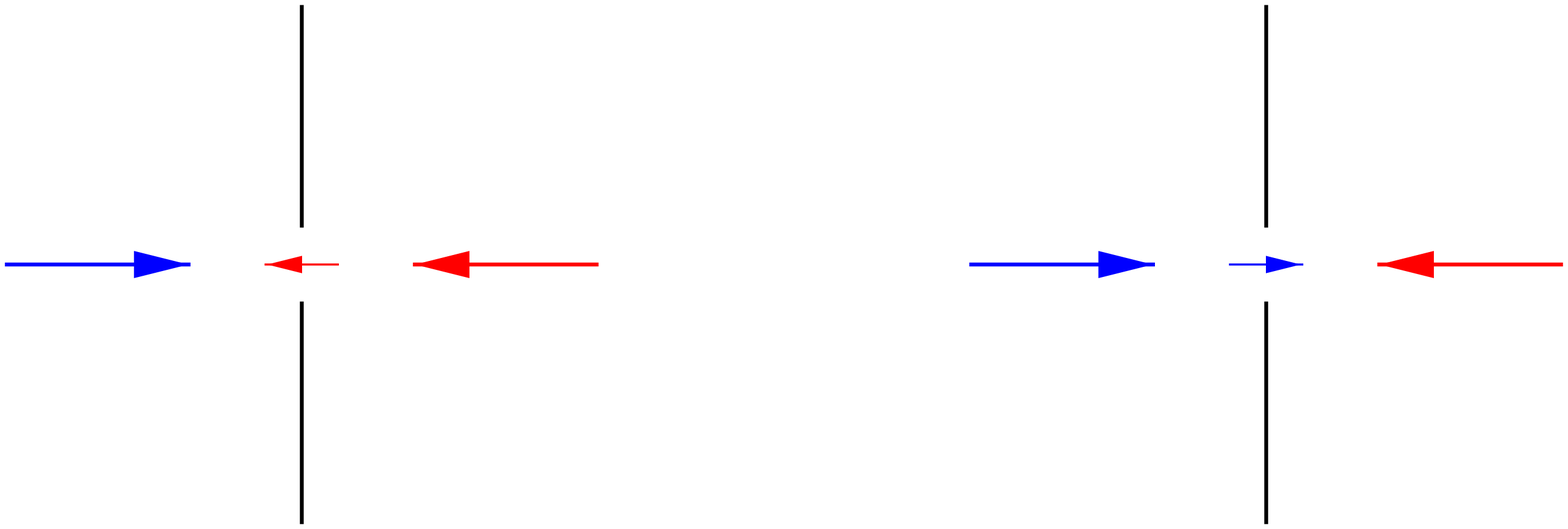}
    \caption{Oscillations of the flow direction at a door with counterflow.}
\label{fig_dooroszi}
  \end{center}
\end{figure}

{\bf Oscillations}: In counterflow at bottlenecks, e.g.\ doors,
oscillatory changes of the direction of motion are observed.
Once a pedestrian is able to pass the bottleneck it becomes easier
for others to follow her/him in the same direction until somebody is
able to pass (e.g.\ through a fluctuation) the bottleneck in the
opposite direction (see Fig.~\ref{fig_dooroszi}).
%%%%%%%%%%%%%%%%%%%%%%%%%%%%%%%%%%%%%%%%%%%%%%%%%%%%%%%%%%%%%%%%%
%%%%%%%%%%%%%%%%%%%%%%%%%%%%%%%%%%%%%%%%%%%%%%%%%%%%%%%%%%%%%%%%%

{\bf Patterns at intersections}: At intersections various collective
patterns of motion can be formed. A typical example are short-lived
roundabouts which make the motion more efficient. Even if these are
connected with small detours the formation of these patterns can
be favourable since they allow for a ``smoother'' motion.

{\bf Panics}: In panic situations, many counter-intuitive phenomena 
can occur. 
In the faster-is-slower effect \cite{panic} a higher desired velocity 
leads to a slower movement of a large crowd. Typical is also
herding behaviour where people just blindly follow others.
Such effects are
extremely important for evacuations in emergency situations.
%In the freezing-by-heating effect \cite{HFV}
%increasing the fluctuations can lead to a more ordered state.
%For a thorough discussion we refer
%to \cite{HePED,panic} and references therein. 

%%%%%%%%%%%%%%%%%%%%%%%%%%%%%%%%%%%%%%%%%%%%%%%%%%%%%%%%%%%%%%%%%%%%%%%%%%%%%%

%%%%%%%%%%%%%%%%%%%%%%%%%%%%%%%%%%%%%%%%%%%%%%%%%%%%%%%%%%%%%%%%%
\subsection{Modelling Pedestrian Dynamics}
%%%%%%%%%%%%%%%%%%%%%%%%%%%%%%%%%%%%%%%%%%%%%%%%%%%%%%%%%%%%%%%%%

Several different types of models have been suggested 
in order to reproduce and understand the phenomena described
in the previous subsection. In addition, practical applications,
e.g.\ in the planning of public buildings like football stadiums,
are of considerable importance. The latter requires models
that can simulate even large crowds efficiently but, at the same
time, are realistic enough to capture the essential aspects of the
dynamics, e.g.\ the observed collective effects.
Therefore it is not surprising that only a few models are able
to achieve this.

A continuum approach that has been very successful in modelling
pedestrian dynamics, is the so-called social force model
(see e.g.\ \cite{helbing,social} and references therein). 
Pedestrians are treated as particles subject to long-ranged forces
induced by the social behaviour of the individuals. The typical structure
of the force between the pedestrian is described by \cite{panic}
\begin{equation}
\mathbf{f}_{ij} = \mathbf{f}_{ij}^{(soc)}+ \mathbf{f}_{ij}^{(phys)}.
\end{equation}
Here $\mathbf{f}_{ij}^{(phys)}$ is a physical force and describes friction and
compression when pedestrians make contact.
$\mathbf{f}_{ij}^{{(soc)}}$ is a repulsive social force modelling the
tendency to keep a certain distance to other individuals. 
Typically it is long-ranged and has the form
\begin{equation}
\mathbf{f}_{ij}^{(soc)}=A_ig_{ij}(\lambda_i,\varphi_{ij})
\exp\left(r_{ij}/\xi_i\right)\mathbf{n}_{ij}
\end{equation}
where $r_{ij}$ is (a suitably defined) distance between the pedestrians
and $\mathbf{n}_{ij}$ a normalized vector pointing from individual $j$
to $i$.  Apart from the interaction strength $A_i$ and the range $\xi_i$
of the force it has also a direction dependence that enters through
the function $g_{ij}$ which depends on a parameter $\lambda_i$ controlling
the anisotropy of interactions and the angle $\varphi_{ij}$ between
the directions of motion.

This idea leads then to equations of motion similar to Newtonian 
mechanics. There are, however, important differences since, e.g., 
in general the third law (``action = reaction'') is not fulfilled by social
forces.

So-called active-walker models \cite{activewalker,trail} have been
used to describe the formation of human or animal trails etc. 
Here the walker leaves a trace by modifying the underground on his path. 
This modification is similar to chemotaxis since it can be regarded
as a stimulus for other pedestrians. Vegetation is destroyed 
by the walker and so it becomes more attractive for others to follow
the same path.

Most cellular automata (CA) models for pedestrian dynamics proposed so 
far are rather simple \cite{fukui,nagatani99,naga00,hubert} and can 
be considered as generalizations of the Biham-Middleton-Levine model 
for city traffic \cite{BML}.
However, these models are not able to reproduce all the collective
effects observed empirically. 

In the following we present the so-called floor field CA model
developed in \cite{ourpaper,ourpaper2,friction}.
It is a CA with stochastic dynamics and in many respects can
be regarded as a two-dimensional version of the ant trail
model of Sec.~\ref{sub-anttraffic}. The basic idea is to model 
interactions between pedestrians as a kind of
virtual chemotaxis. Like an ant on a ant trail any {\em moving}
pedestrian creates a {\em virtual} trace that influences
the motion of other pedestrians by enhancing the probability
of motion in the same direction.
In this way long-ranged spatial interactions are 
translated into local interactions, but with ``memory''. This
reduces the number of interaction terms considerably (from $O(N^2)$ to
$O(N)$ for crowds of $N$ people) and allows
for a much more efficient implementation on a computer.

The idea of a virtual trace can be generalized to a so-called {\em floor
field}. This floor field includes the virtual trace created by the
pedestrians as well as a static component which does not change with
time. The latter allows to model e.g.\ preferred areas, walls and other 
obstacles. The pedestrians then react to both types of floor fields. 
The `particles' in the model have very little ``intelligence'' and 
the formation of complex structures and collective effects is solely
achieved through self-organization.  
No detailed assumptions about the human behaviour are necessary.

As already emphasized, it is similar to a 2-dimensional variant of 
the ant trail model. We have a hard-core exclusion so that each cell 
can be occupied at most by one pedestrian. In contrast to pheromone
field in the ant trail model, the floor fields are virtual and, 
therefore, not restricted by hard-core exclusion. Here we do not give 
a complete definition of the model here which can be found in 
\cite{ourpaper,ourpaper2,friction}.

These basic principles are already sufficient to reproduce the effects
described in Sec.~\ref{sec_coll}, such as lane formation in a corridor, 
herding and oscillations at a bottleneck \cite{ourpaper,ourpaper2}. 
In addition, the model can also be used very efficiently for the simulation
of emergency situations \cite{egress}.

\section{Conclusion} Aesthetically beautiful patterns are formed by 
aggregates of living organisms. Such patterns are formed by organisms 
as simple as uni-cellular bacteria, by social insects like ants and 
termites as well as by more complex vertebrates like birds and fish. 
All the patterns of our interest are transient in nature. Interesting 
transient patterns emerge also in human societies. During time 
intervals that are short compared to the lifetime of a pattern formed 
by the terrestrial locomotion of organisms, the collective movements 
on linear segments often appear similar to vehicular traffic. In 
this article we have presented a critical review of the theoretical 
works, particularly those published over the last decade, focussing  
almost exclusively on the agent-based models formulated in the spirit 
of the classical Lagrangian approach. These include the Langevin-like 
stochastic differential equations, where dynamics is formulated in 
continuous space-time to describe the trajectories of the individual 
organisms. Another class of agent-based discrete models are developed  
using the language of cellular-automata where the dynamics is formulated 
in terms of update rules. Direct comparison with controlled experiments 
have been possible mostly in case of micro-organisms and small insects. 
The challenge is to unveil the mystery of these transient tapestry and 
traffic of life.

%%%%%%%%%%%%%%%%%%%%%%%%%%%%%%   END TEXT   %%%%%%%%%%%%%%%%%%%%%%%%%%%%

\noindent {\bf Acknowledgements} \\

\noindent DC thanks Dietrich Stauffer for hospitality in K\"oln during 
a visit, supported by a joint Indo-German research project funded by 
DFG, when some of the works reported here were initiated. We also 
thank Alexander John for useful discussions.

\end{document}